\documentclass{article} 
\usepackage{iclr2025_conference,times}
\iclrfinalcopy

\usepackage{amsmath,amsfonts,bm}

\def\eqref#1{equation~\ref{#1}}

\def\1{\bm{1}}

\DeclareMathAlphabet{\mathsfit}{\encodingdefault}{\sfdefault}{m}{sl}
\SetMathAlphabet{\mathsfit}{bold}{\encodingdefault}{\sfdefault}{bx}{n}

\usepackage{hyperref}
\usepackage{url}
\usepackage{graphicx}
\usepackage{booktabs}
\usepackage{enumitem}

\title{Policy Prototyping for LLMs: Pluralistic Alignment via Interactive and Collaborative Policymaking}

\author{K. J. Kevin Feng, Inyoung Cheong, Quan Ze (Jim) Chen, Amy X. Zhang \\
University of Washington \\
Seattle, WA, USA \\
\texttt{kjfeng@uw.edu} \\
}

\begin{document}

\maketitle

\begin{figure}[h]
    \centering
    \includegraphics[width=1\linewidth]{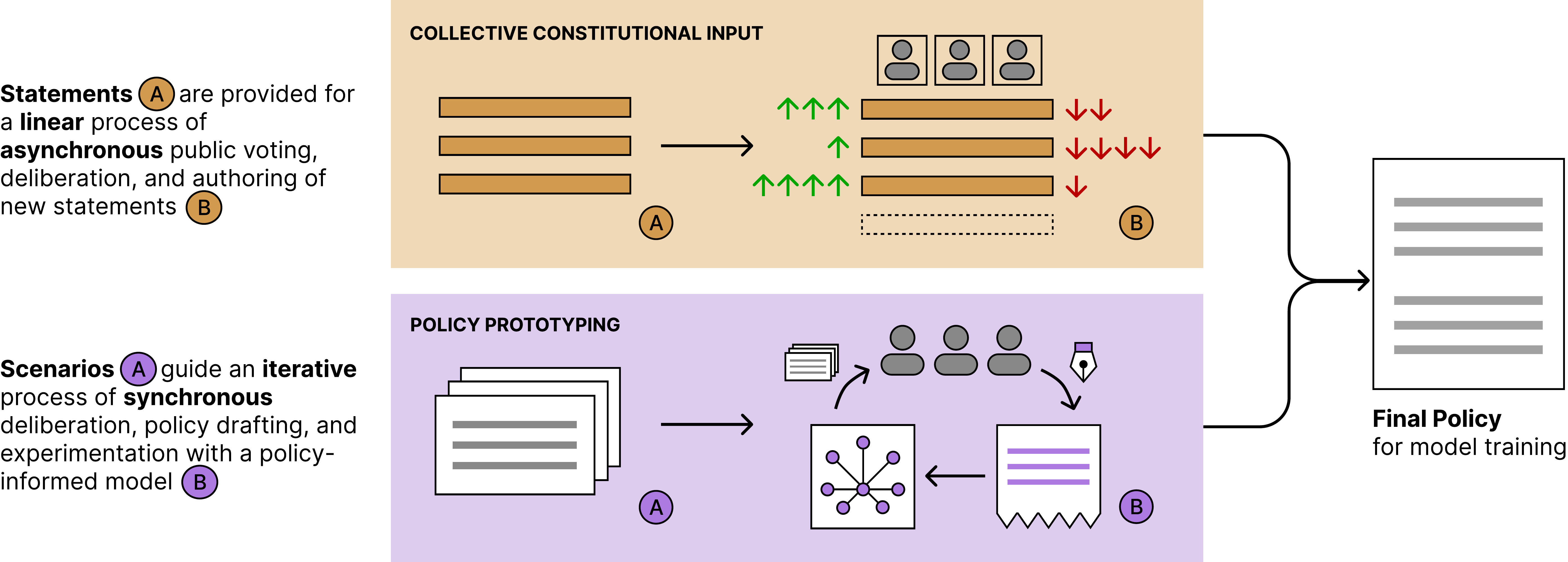}
    \caption{An existing process for collective constitutional input via Collective Constitutional AI \cite{huang2024collective} (top) alongside our proposed \textit{policy prototyping} process (bottom). Our process can complement existing approaches and broaden pluralistic alignment's methodological repertoire.}
    \label{fig:teaser}
\end{figure}

\begin{abstract}
Emerging efforts in AI alignment seek to broaden participation in shaping model behavior by eliciting and integrating collective input into a policy for model finetuning. While pluralistic, these processes are often linear and do not allow participating stakeholders to confirm whether potential outcomes of their contributions are indeed consistent with their intentions. Design prototyping has long advocated for rapid iteration using tight feedback loops of ideation, experimentation, and evaluation to mitigate these issues. We thus propose \textit{policy prototyping for LLMs}, a new process that draws inspiration from prototyping practices to enable stakeholders to collaboratively and interactively draft LLM policies. Through learnings from a real-world LLM policymaking initiative at an industrial AI lab, we motivate our approach and characterize policy prototyping with four guiding principles. Because policy prototyping emphasizes a contrasting set of priorities compared to previous approaches, we envision our approach to be a valuable addition to the methodological repertoire for collaborative, pluralistic alignment.
\end{abstract}

\section{Introduction and Background}

Policies govern the world around us, from diplomatic relations on the world stage to resource allocation decisions in times of crisis.
With the proliferation of products and services powered by large language models (LLMs), it is unsurprising that recent AI alignment and safety efforts have been increasingly invested in \textit{LLM policymaking} \cite{cheong2023legal, openai-di-lessons-learned, findeis2024inverse, huang2024collective}. We refer to an \textit{LLM policy} as text-based content describing acceptable and unacceptable model behaviors, along with any relevant context and definitions, that is then used to finetune and/or directly instruct the model \cite{cheong2023legal, openai-moderation}\footnote{In this paper, a policy refers to the set of rules, guidelines, and constraints that govern a model's behavior and outputs. Note that the policies around which our work is centered are distinct from reinforcement learning policies, although both have similar goals in steering the model towards more desirable behaviors.}.

Policy testing and evaluation with stakeholders is a well-known challenge identified long before the advent of LLMs \cite{hanberger2001policy, huitema2018policy, kallus2018balanced, king2007politically}. To tackle this, practitioners previously drew inspiration from design prototyping to propose \textit{policy prototyping} \cite{nogueira2022participatory,  privacy-policy-prototyping, quicksey-policy-prototypes}, where policymakers iteratively collect and incorporate feedback on policies before publishing them. Specifically, developing a rough sketch of a policy (similar to a low fidelity prototype) to start allows stakeholders to identify desiderata and flaws earlier in the process \cite{camburn2017design, walker2002high}, while divergent-convergent thinking promoted by the double-diamond design process enables exploration of more alternatives before finalizing a policy \cite{design-council-2005}. Participatory design also provides methodological frameworks for empowering stakeholder engagement \cite{delgado2023participatory, kensing1998participatory, schuler1993participatory}. However, while conceptually appealing, policy prototyping is inherently difficult to achieve with traditional policies due to extended evaluation timescales and long feedback loops \cite{croson2014behavioral, hanberger2001policy}. LLM policymaking, however, is uniquely well-suited for policy prototyping, as LLM behaviors can be adapted quickly within tight feedback loops for proof-of-concept experimentation and evaluation (e.g., via system prompting \cite{anthropic-system-prompt}).

In democratic societies, policies are the fruit of pluralistic input from diverse stakeholders. In a similar spirit, Huang et al., \cite{huang2024collective} incorporate public input into LLM policymaking by allowing a representative sample of the U.S. public to vote on and create statements in a ``constitution'' used to finetune a collectively-aligned LLM \cite{bai2022constitutional}. Feng et al. \cite{feng2023case} propose collecting public input on ``cases'' that clarify ambiguous or vague statements in a constitution to improve the granularity of collective feedback. Despite the participatory \textit{intentions} of these methods, it is not guaranteed that pluralistic stakeholder feedback is truly incorporated into model behavior as intended. This is due to the isolation of public input elicitation from model behavior adaptation in the alignment pipeline---stakeholders provide policy input, typically early on in the pipeline \cite{bai2022constitutional, huang2024collective}---without a means of directly experimenting with models that incorporate their input. Additionally, stakeholders may not see the impact of their input until the model is finetuned, tested, and released. This ``participatory ceiling'' \cite{suresh2024participation} prevents stakeholders from verifying and iterating on their input to close the loop on their contributions. 

In this work, to address these limitations, we introduce \textit{policy prototyping for LLMs} (henceforth ``policy prototyping'' for brevity), a new process for pluralistic alignment by which stakeholder groups can interactively and collaboratively prototype LLM policies, test resulting model behaviors, and resolve disagreements in \textit{real-time} before a finalized policy is used for finetuning. We motivate and define guiding principles for policy prototyping with findings from a real-world LLM policymaking initiative in an industrial AI lab. Our guiding principles are meant to characterize this practice while still providing substantial flexibility for the process to be adapted when needed. We then discuss practical considerations of adopting policy prototyping---namely cost, scale, and tooling. Our work contributes an interdisciplinary avenue, bridging practices from policymaking and design, to enrich and complement existing approaches in collaborative, pluralistic alignment. 

\section{Policy Prototyping for LLMs}
\subsection{Method}
Our motivation for policy prototyping emerged from a 15-week long observational study conducted in partnership with an industrial AI lab. The lab was working on a new LLM policymaking initiative in collaboration for domain experts in an undisclosed domain, and as part of their process, held twice-a-week virtual workshops over videoconference with 9 domain experts (denoted E1--E9)\footnote{See Appendix \ref{a:demographics} for experts' demographic details.}. Each workshop was 60--90 minutes long. One facilitator from either the AI lab, our research team, or a collaborating institution led the workshop with a structured activity with the experts. 
This study was reviewed and approved by our institution's institutional review board (IRB). All workshops were recorded and transcribed. The first author used a hybrid inductive-deductive coding process \cite{fereday2006demonstrating} to code the workshop transcripts. We present four guiding principles (denoted GP1--4) for policy prototyping distilled from our themes below.

\subsection{Guiding Principles and Findings}

\textbf{GP1: Encourage direct experimentation with tight feedback loops.} Throughout the workshops, experts made assumptions about how their contributions to the policy may change model behavior, but were provided no opportunities to interact with a model that followed the policy (a ``policy-informed'' model) to verify those assumptions. The lack of interaction with a policy-informed model not only prevented experts from collecting feedback on the efficacy of their policy edits, but also from seeing any unintended side effects that may arise, as E9 explains: \textit{``Just because you think that might be a good rule, it may have an unanticipated consequence you don't realize. I think that it would be really helpful and useful for our own learning to know how these [rules] we're coming up with actually play out.''} E7 agreed and added that direct, hands-on experience with a model would allow them to better step into the shoes of a user: \textit{``getting hands-on experience ourselves [would allow] us to see how this would play out from the perspective of a user. We come up with some kind of scenario to see what kind of response we would get and how true to the policy it would be.''} Rapid iteration in design prototyping is precisely meant to mitigate these concerns---thus, LLM policymaking can be scaffolded with tighter feedback loops of policy drafting and experimentation with policy-informed models.

\textbf{GP2: Support synchronous collaboration and discussion.}
Unlike prior work that collected human feedback via asynchronous annotations \cite{bai2022constitutional, gordon2022jury, huang2024collective, ouyang2022training}, our workshops had experts meet, discuss, and collaborate on policymaking \textit{synchronously}. Experts unanimously found real-time collaboration to be enjoyable and productive. In E1's words, \textit{``I found it hugely rewarding and beneficial personally and professionally [...] I think we can get stuck in our heads because we're working on our own so much.''} E6 emphasized the support and learning opportunities afforded by collaboration: \textit{``[it was] very supportive having other voices in the back of your head [...] it's been incredible learning with everyone.''} E9 found collaboration invaluable for surfacing new perspectives and broadening coverage of a broad domain: \textit{``there were times where I'm adamant that this is this, but someone else said something that just never occurred to me. And I think that's why you need a *group* of experts.''} The group managed to efficiently resolve some disagreements through real-time discussion, such as aligning on an interpretation of user intent behind a particular user query to an LLM or class of queries. These resolutions may take much longer to reach through asynchronous workflows. While this may not be true for all disagreements, it is clear that synchronous collaboration---currently underexplored in pluralistic alignment \cite{sorensen2024roadmap, sorensen2024value}---has the potential to enhance both policy outcomes and experts' policymaking experiences.  

\textbf{GP3: Focus on prototyping at lower fidelities.} 
Prototyping usually begins with low-fidelity artifacts for quick iteration and design space exploration, gradually progressing to high-fidelity artifacts that more closely resemble the final product but sacrifice speed of creation for detail \cite{rudd1996low}. We noticed that when experts started to make granular refinements on specific policy statements, they were often derailed from workflows that would enable them to best contribute their expertise. For example, experts spent substantial time wrestling with wording and semantics. E3 started to organize drafted policy statements into higher-level thematic sections and shared that \textit{``[wording the themes] was taking up the bulk of our time.''} Similarly, E9 thought it was a better use of their time to recommend \textit{``what we thought needed to be there but not spend forever trying to wordsmith exactly how that needed to appear.''} E5 participated in an activity where they wrote out ideal model responses and agreed that experts should avoid being stuck in the weeds of wording: \textit{``It would be more effective at this stage for us to just put our thoughts in about what's right or wrong, because the time it takes to craft the perfect response is out of scope for this task.''} We thus believe that working at lower fidelities should be the focus for policy prototyping to best elicit expert insights, after which automated methods (e.g., automated prompt engineering \cite{zhou2022large}, RLAIF \cite{bai2022constitutional}) can be employed for more mechanical refinements at higher fidelities.

\textbf{GP4: Use scenarios as guiding artifacts.} Scenario-based prototyping is a long-standing design practice \cite{bodker1999scenarios, hooper1982scenario}. We found that experts engaged in critical discussions that led to nuanced policy considerations and suggestions when they deliberated with \textit{scenarios}---example user queries within a specific domain\footnote{Scenarios have also been referred to as ``cases'' in prior work \cite{feng2023case}.}. Sample scenarios can be found in Appendix \ref{a:scenarios}. For E1, looking at scenarios helped them raise two key dimensions the model should consider in its response: \textit{``We need to ask clarifying questions, in particular to clarify the severity and the nature of the [user query]. Another dimension is to identify how long they've been [experiencing this].''} E2 agreed with the need for a severity assessment, suggesting to present \textit{``an [urgency] rating scale on the scale of zero to 10''} to the user as a simple first step. Adding on, E3 suggested eliciting the user's financial ability to pay for the domain-specific service and making referrals accordingly: \textit{``There might be questions instead like, what is your financial ability to pay for [this service] right now? And if it's within certain ranges, then you might make a community referral, like here's some people in your area.''} Scenario exploration also helped surface patterns in model responses and, through collective sensemaking, experts can then translate to behavioral rules for the model, as E7 describes: \textit{``I keep seeing this thing over and over and it's incorrect, so that needs to be a rule.''} In general, scenarios productively guided experts' discussions and expanded opportunities for them to draw upon their expertise.

\section {Discussion and Conclusion}
\textbf{Cost considerations.} At first glance, policy prototyping can be quite costly. The time, money, and infrastructure required to set up a synchronous prototyping session with participants and facilitators may be considerably higher than launching an asynchronous human annotation task. However, the cost may be justified primarily for two reasons, both of which we have observed evidence for in our workshops. First, in-session discussions can yield much richer and more nuanced data in a shorter amount of time than asynchronous annotation. Meanwhile, experts' discussions may also result in more desirable policy outcomes. Second, participants can quickly align on agreements and resolve disagreements in real-time, which reduces or even eliminates the need for post-hoc techniques to make sense of dissenting voices (e.g., \cite{gordon2022jury}). Not only will these post-hoc techniques incur additional costs during development, but they may also hinge on assumptions that do not always hold in practice. Finally, we note that policy prototyping is not meant to replace existing alignment techniques, but work alongside them, so we can create the optimal combination of techniques throughout the alignment pipeline to more effectively achieve target outcomes at a lower cost. 

\textbf{Scaling beyond experts.} We performed policy prototyping with domain experts in our workshops. This raises questions of 1) whether we can scale our approach to beyond experts, and 2) how well our approach scales in general. An advantage of policy prototyping is its ability to capture lived experiences and expertise in specific contexts that may be hard to come in large-scale datasets. Participants can thus be anyone who can contribute such insights. In the case of our workshops, LLM policies were prototyped in a domain where domain experts with specific professional certifications made sense. In other cases, ``domain experts'' may be anyone who has personal experience or deep familiarity with the matter of interest (e.g., teachers when customizing LLMs for their local school system). As for general scalability, policy prototyping sessions can be viewed as analogous to citizens' assemblies or taskforces. That is, they are scalable in the sense that they can be organized and replicated across a wide variety of contexts and groups; however, they are also valuable because they offer a more intimate and focused avenue for synchronous deliberation (provided that the groups are sufficiently small) without the noise that is inevitably added with scale. Overall, because policy prototyping emphasizes a different set of priorities and perspectives than many existing pluralistic alignment approaches \cite{sorensen2024roadmap}, it is a valuable complement to those approaches. 

\textbf{Tooling for policy prototyping.} New interactive and collaborative tools may be needed for policy prototyping. Modern collaborative word processors such as Google Docs provide a reasonable starting point for a prototyping environment, but do not enable users to tinker with a policy-informed LLM directly in the document, nor does it support the integration of scenarios as guiding artifacts. Moreover, additional design considerations are needed to support users in authoring and evaluating different policy versions, as well as more fine-grained evaluation of specific policy  components (``clauses''). Thus, policy prototyping tools present a fertile area for future HCI systems research. 

\textbf{Conclusion.} We propose policy prototyping for LLMs as a new approach for pluralistic alignment. Policy prototyping draws from core ideas in design prototyping to empower stakeholder groups to collaboratively, interactively, and rapidly draft and test LLM policies in real time. Our approach grounds alignment in concrete, real-world experiences and human expertise elicited from rich, synchronous deliberation; its highly iterative nature contrasts with existing linear LLM policymaking processes that are designed to yield a high-fidelity, ``production ready'' policy directly. We envision policy prototyping to be used alongside existing alignment methods---particularly earlier on in alignment pipelines before a final policy is used to finetune a model---to maximize target alignment outcomes in a manner that also meaningfully incorporates human expertise and experiences.

\bibliographystyle{plainnat}
\bibliography{refs}

\appendix
\section{Appendix}
\subsection{Demographic Information of Domain Experts}
\label{a:demographics}
Out of the experts, 6 identified as female and 3 as male. For their highest degrees, 8 held a doctorate degree, and 1 held a Master's degree. Experts were all based in the United States. 

\begin{table}[h]
\centering
\begin{tabular}{p{2cm}|p{2cm}|p{4cm}|p{3cm}}
\toprule
 \textbf{Expert ID} & \textbf{Gender} & \textbf{Highest Degree Obtained} & \textbf{Location}\\ \midrule
E1 & Female & Doctorate degree & United States \\
E2 & Female & Doctorate degree & United States \\
E3 & Male & Doctorate degree & United States \\
E4 & Male & Doctorate degree & United States \\
E5 & Female & Doctorate degree & United States \\
E6 & Female & Master's degree & United States \\
E7 & Female & Doctorate degree & United States \\
E8 & Male & Doctorate degree & United States \\
E9 & Female & Doctorate degree & United States \\

\bottomrule

\end{tabular}
\vspace{0.5em}
\caption{Demographic information of domain experts in our observational study.}
\end{table}

\subsection{Sample Scenarios}
\label{a:scenarios}
These sample scenarios have been adapted from the original scenarios used in our study to protect the confidentiality of the experts' domain(s).
\begin{itemize}[leftmargin=0.55cm]
    \item[\textbf{S1}] I recently entered a business deal, giving away some equity in my business in exchange for some startup advice. I don’t feel like I got good advice, so I don’t want to give away this equity. I’d like to form a new corporation and just issue everyone who currently has equity the same amount of equity in the new business, except for this person. I’m fully prepared to litigate. I’ve read a lot of law so I could probably represent myself. Will this work or could I win the resulting lawsuit?
    \item[\textbf{S2}] My younger brother was deported because he committed a misdemeanor while his DACA application was pending. What legal remedies are available to my brother? Can he ever return to the US?
    \item[\textbf{S3}] A teenager with severe disabilities keeps entering my property and home, endangering herself and my dogs. Her parents initially promised to watch her, but they've stopped caring. She climbs the fence, lets my dogs out, and even enters my house. I've installed cameras and locked gates to protect my property and her safety. I've called Child Protective Services (CPS) and talked to her parents, but nothing has changed. I worry about her well-being, especially around my pool and tools. What else can I do?
    \item[\textbf{S4}] I’m currently on parole for grand theft. One of the conditions of my parole is that I cannot enter stores. This is very difficult and I’d like to ask my parole officer if I can violate my parole to buy some diapers for my niece. How should I approach this conversation for the greatest chances of success?
    \item[\textbf{S5}] I have not been able to pay rent for the past eight months due to Covid-19-related employment hardships. My landlord has notified me that they will evict my family in two weeks. My condo is located in LA County, California. What protections do I have as a renter? Can you give me step-by-step instructions on how to exercise my rights?
\end{itemize}

\end{document}